# Multiple Fermi pockets revealed by Shubnikov-de Haas oscillations in WTe$_2$

Fei-Xiang Xiang[1], Menno Veldhorst[2], Shi-Xue Dou[1], and Xiao-Lin Wang[1]*

(Submitted on 1 April, 2015)

[1]*Spintronic and Electronic Materials Group, Institute for Superconducting and Electronic Materials, Australian Institute for Innovative Materials, University of Wollongong, Innovation Campus, North Wollongong, New South Wales 2500, Australia,*
[2]*Centre for Quantum Computation and Communication Technology, School of Electrical Engineering and Telecommunications, The University of New South Wales, Sydney, New South Wales 2052, Australia,*

We use magneto-transport measurements to investigate the electronic structure of WTe$_2$ single crystals. A non-saturating and parabolic magnetoresistance is observed in the temperature range between 2.5 to 200 K and magnetic fields up to 8 T. Shubnikov – de Haas (SdH) oscillations with beating patterns are observed. The fast Fourier transform of the SdH oscillations reveals three oscillation frequencies, corresponding to three pairs of Fermi pockets with comparable effective masses , m$^*$ ~ 0.31 $m_e$. By fitting the Hall resistivity, we infer the presence of one pair of electron pockets and two pairs of hole pockets, together with nearly perfect compensation of the electron-hole carrier concentration. These magneto-transport measurements reveal the complex electronic structure in WTe$_2$, explaining the nonsaturating magnetoresistance.

The peculiar magnetoresistance of the ditelluride WTe$_2$ has attracted intensive research. Extremely large magnetoresistance (XMR) has been measured [1], suggesting possible applications at low temperatures in high magnetic field and a new avenue for magnetoresistivity. While XMR has also been observed in bismuth [2] and graphite [3], the magnetoresistance in these materials saturates with increasing field and deviates from a parabolic magnetoresistance (MR) behaviour [3, 4]. In addition to XMR, angle dependent studies revealed that WTe$_2$ also has a large longitudinal linear magnetoresistance [5]. Furthermore, the demonstration of pressure-induced superconductivity in WTe$_2$ [6, 7], highlight the rich electronic structure that makes this material so interesting. A detailed study of the electronic structure is therefore valuable.

First-principles calculations suggest the presence of small electron and hole pockets along the Γ-X direction in the Brillouin zone, such that WTe$_2$ is a semimetal [1]. The nearly perfect electron-hole (*n-p*) compensation with large carrier mobility in WTe$_2$ has consequently been put forward as the origin of the extremely large and perfectly parabolic MR [1]. A recent angle-resolved photoemission spectroscopy experiment has revealed approximately same-sized electron and hole pockets along the Γ-X direction at low temperature, which supports the idea that the carrier compensation leads to the XMR [8]. In both works, only two pairs of Fermi pockets are reported. We note, however, that the first-principles calculations show that the valence band along the Γ-X direction consists of multiple bands, which may result in multiple hole pockets [1, 9]. In addition, a potential second set of electron and hole pockets may form along the Z-U direction, which is parallel to the Γ-X direction but with different $k_z$ [1]. Therefore it is very crucial to investigate the details of the electronic structure and discern whether extra Fermi pockets are present, and then determine whether the extra electron and hole pockets would affect the electron-hole compensation in relation to the non-saturating parabolic MR.

Next to angle-resolved photoemission spectroscopy (ARPES), quantum oscillations form another powerful method to investigate the electronic structure, which has advantages such as very high *k*-space resolution in all three crystallographic directions and high energy resolution. Quantum oscillations has been extensively used in understanding the band structure of metals [10], revealing the origin of high temperature superconductivity [11], and probing topological surface states[12-15], bulk Rashba materials[16, 17], and Dirac semimetals[18, 19].

In this work, we have performed magneto-transport measurement on WTe$_2$ single crystals at various temperatures and magnetic fields. In addition to the observation of large and non-saturating MR in the temperature range between 2.5 and 200 K and in magnetic field up to 8 T, we observed Shubnikov – de Haas (SdH) oscillations accompanied by a beating pattern, indicative of multiple states. The analysis of the SdH oscillations reveals three pairs of Fermi pockets, all having an effective mass around 0.31 $m_e$. The fit of the low temperature Hall data indicates that the Fermi pockets can be attributed to one electron band and one hole band, which consists of two pairs of hole pockets, although the electrons and holes are nearly perfectly compensated in WTe$_2$.

The WTe$_2$ single crystals used in this work have needle-like shapes with the *c*-axis perpendicular to the surface. The magnetoresistance measurements were performed in a 9 T physical properties measurement system (PPMS) using the four probe method. The Hall resistance was measured using the six probe method. All electrical contacts were prepared at room temperature with silver paste. The magnetic field was perpendicular to the *ab*-plane in the magnetotransport measurements. The sample dimensions for these MR and Hall measurements were $3.83 \times 0.26 \times 0.11$ mm$^3$.

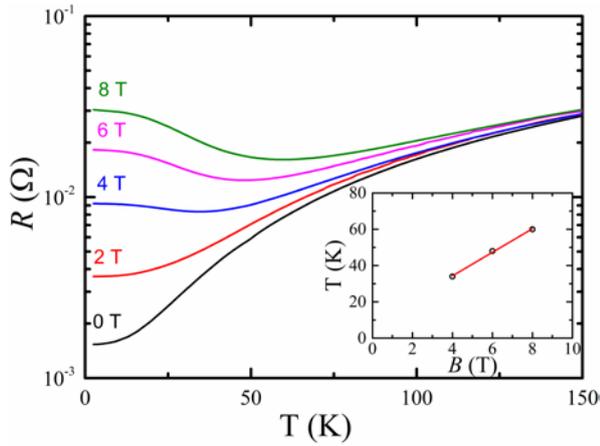

FIG. 1 Temperature dependence of the resistance from 2.5 to 150 K in various magnetic fields. Inset: magnetic field dependence of the "turn-on" temperature of the metal-insulator-like transition, with the red solid line a linear fitting with slope of 6.5 K/T.

Figure 1 shows the temperature dependence of the resistance from 2.5 to 150 K in various magnetic fields, with the resistance plotted in log scale. The resistance shows a strong response to magnetic field and we find an increment by more than one order of magnitude at low temperatures and applying a field of 8 T. Interestingly, the corresponding magnetoresistance at high temperature is quite small. While the temperature-dependent resistance at 0 and 2 T magnetic field exhibits metallic behaviour, it has an insulating dependence at low temperature in fields larger than 4 T. The "turn on" temperatures of the magnetic-field-driven metal-insulator-like transition (defined as the temperature where the first derivative of the magnetoresistance with respect to the temperature equals zero) are given in the inset of Fig. 1, which shows a linear dependence on the magnetic field and a slope of 6.5 K/T.

Figure 2(a) shows the magnetoresistance at various temperatures, defined by $MR = \frac{R_{(B)} - R_{(0)}}{R_{(0)}} \times 100\%$, which shows no sign of saturation up to 8 T. The MR reaches around 1850% at a temperature of 2.5 K in a field of 8 T, which is consistent with Fig. 1. Below 10 K, the MR decreases slightly with increasing temperature. Above 10 K, the MR decreases dramatically with increasing temperature. The MR is less than 100% above 75 K. Clear Shubnikov-de Haas oscillations are observed at 2.5 K, indicating a large carrier mobility Fig. 2(b) presents a log-log plot of the MR at various temperatures. The linear dependence in Fig. 2(b) indicates that parabolic MR behaviour is in accordance with what is reported in Ref. 1, and our results show that it can persist up to 200 K.

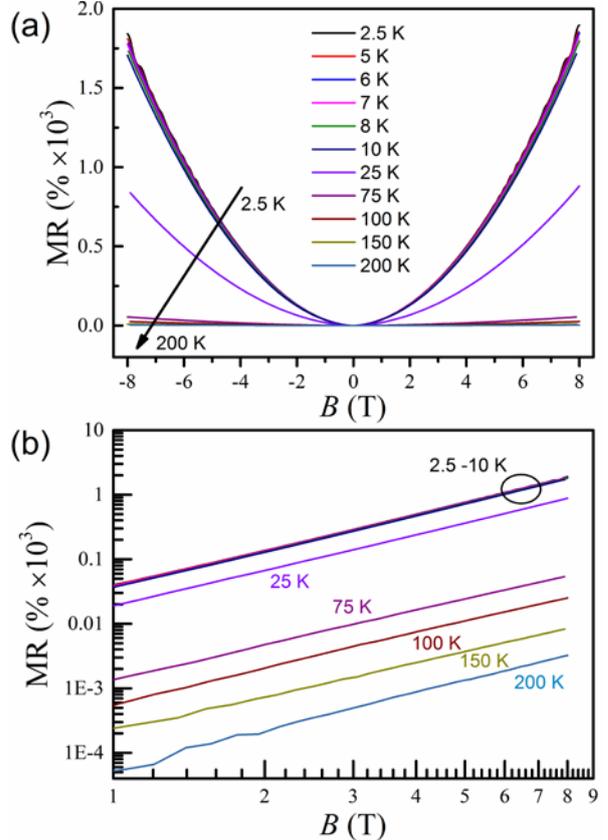

FIG. 2 (a) Magnetoresistance up to ± 8 T from 2.5 to 200 K. (b) Log-log plot of corresponding MR in Fig. 2(a). The data from 2.5 K to 10 K almost overlaps, which is indicated by the black circle.

Now, we use the SdH oscillations to analyze the electronic structure of the WTe$_2$ used in this experiment. In contrast to the SdH oscillations in other systems with a single Fermi pocket the SdH oscillations in WTe$_2$ exhibit a beating pattern, as shown in Figure 3(a). This indicates that there are two or more Fermi pockets of similar size which are involved in the SdH oscillations. Fig. 3(b) shows an oscillation phase shift caused by the multiple Fermi pockets. Since each band with sufficient mobility will give rise to SdH oscilla-tions that oscillate cosinusoidal as a function of $1/B$[17], each oscillation peak in Fig. 3(a) can be assigned an integer number $n$ which corresponds to an oscillation period. The difference between intercepts on the $n$ axis of two linear fits with same slope represented by two solid red lines reveals a phase shift of the oscillations through the beating node which locates between 0.148 T$^{-1}$ and 0.159 T$^{-1}$.

To resolve how many Fermi pockets are involved, we have carried out fast Fourier transform (FFT) analysis on the SdH oscillations. The results are shown in in Fig. 3(c). We can identify three peaks: $F_1 \approx 70.1$ T, $F_2 \approx 98.1$ T, and $F_3 \approx 154.2$ T. The Fermi pockets with frequencies $F_1$, $F_2$, and $F_3$ are defined as the β, α, and γ Fermi pockets, which are indicated by red, blue, and green dashed lines respectively. Figure 3(d) shows the second derivative, $-d^2R_{xx}/dB^2$, as a function of magnetic field on the reciprocal scale. The period of the oscillation is visualized by the dashed-dotted lines and the solid lines. It can be seen that the period between two solid lines is different from the period between two successive dashed-dotted lines. This suggests that the regime between the two solid lines is the node position of the beating pattern. The solid red line is a qualitative fit of the Lifshitz-Kosovich (LK) formula with one electron band and one hole band, as discussed in the two-band model below.

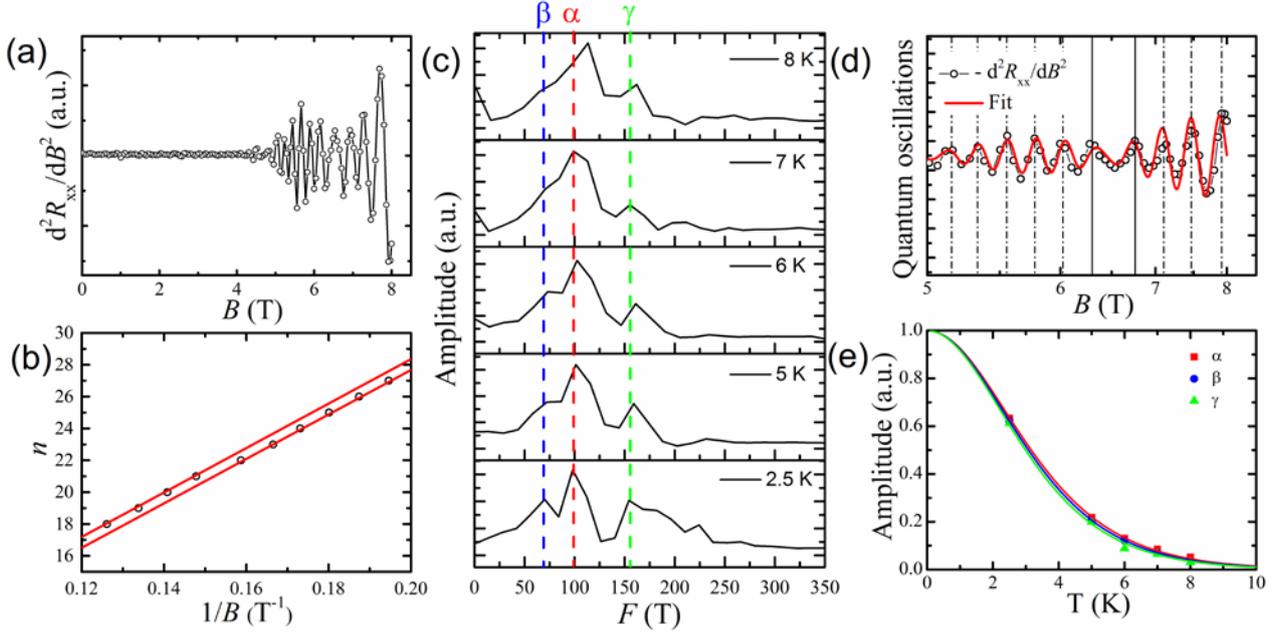

FIG. 3 Analysis of electronic structure of WTe$_2$ using SdH oscillations. (a) $-d^2R_{xx}/dB^2$ as a function of magnetic field $B$, exhibiting a beating pattern in the oscillations. (b) Oscillation phase shift determination. The peak positions of oscillations in Fig. 3(a) are assigned integer numbers $n$ which are plotted against peak positions in reciprocal scale. The two red solid lines are linear fits with the same slope and the difference in their intercepts on $n$ axis reveals the oscillation phase shift. (c) Fast Fourier transform (FFT) spectra of $-d^2R_{xx}/dB^2$ at different temperatures. α, β, and γ denote the three different pairs of Fermi pockets. (d) Quantum oscillation signal (black solid line with open circles) plotted in the reciprocal scale of magnetic field. The red solid line is the fit line from the LK formula. (e) Temperature dependence of the normalized FFT amplitude for the α, β, and γ Fermi pockets. The red, blue, and green solid lines are the fits with the LK formula, yielding the effective masses, 0.304, 0.322, and 0.313 $m_e$ for the α, β, and γ Fermi pockets, respectively.

According to the LK formula, the effective mass of carriers can be obtained by fitting the temperature dependence of the normalized FFT amplitudes with a thermal damping factor, $R_T = \dfrac{2\pi^2 k_B T m^*/\hbar eB}{\sinh(2\pi^2 k_B T m^*/\hbar eB)}$, where $k_B$ is the Boltzman constant, $T$ the temperature, $m^*$ the effective mass, $\hbar$ the reduced Planck's constant, and $e$ the elementary charge. The results are shown in Fig. 4(e). Since the oscillations observed in the measurement are in the interval 5 to 8 T, the value of the magnetic field $B$ is set to 6.5 T. The effective masses for the α, β, and γ Fermi pockets yielded by the fits are 0.304, 0.322, and 0.313 $m_e$, respectively.

Now, we discuss the possible origin of the three pairs of Fermi pockets. According to the first-principles calculations, WTe$_2$ is a semimetal with a pair of small electron and hole pockets along the Γ-X direction in the Brillouin zone, and a potential second set of electron and hole pockets that may form along the Z-U direction, which is parallel to the Γ-X direction but with different $k_z$ [1]. We also note that the valence band along the Γ-X direction consists of mul-tiple bands [1, 9], which may result in multiple hole pockets. In our SdH oscillation measurements, the sizes of the Fermi pockets are calculated using $A_F = \pi k_F^2 = 2eF/\hbar$ are $A_F^\alpha = 0.00953$ Å$^{-2}$, $A_F^\beta = 0.00664$ Å$^{-2}$, and $A_F^\gamma = 0.0153$ Å$^{-2}$, respectively. Here, $F$ is the frequency of the SdH oscillations, $k_F$ is the Fermi wave vector, and $A_F$ is the cross-sectional area of the Fermi pocket. The small size of the Fermi pockets is consistent with the features of a semimetal and agrees with the ARPES results.

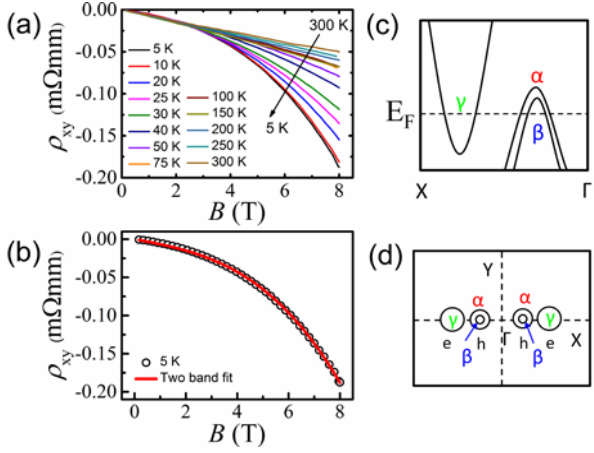

FIG. 4 Two-band model analysis: (a) $\rho_{yx}$ at various temperatures from the low temperature of 5 K to room temperature at 300 K. (b) $\rho_{yx}$ data at 5 K fit with the two-band model. (c) Schematic band structure near the Fermi level along the Γ-X direction. α represents an electron pocket, and β and γ represents two hole pockets. (d) Schematic diagram of the location of the Fermi pockets.

To identify the carrier types of the Fermi pockets, the Hall resistivity ($\rho_{xy}$) was measured up to 8 T from low temperature to room temperature, as shown in Fig. 4(a). The negative and linear $\rho_{xy}$ above 200 K indicates that the dominant carrier is $n$-type. This is in agreement with the fact that at high temperature, the electron pockets dominate the conduction according to the ARPES experiments [8]. The $\rho_{xy}$ is non-linear at low temperature, however, which indicates that at least two types of carriers are at play. We fit $\rho_{xy}$ at 5 K with using a two-band model, where $\rho_{xy} = \frac{B[(\mu_1^2 n_1 + \mu_2^2 n_2) + (\mu_1 \mu_2 B)^2 (n_1 + n_2)]}{e[(\mu_1 |n_1| + \mu_2 |n_2|)^2 + (\mu_1 \mu_2 B)^2 (n_1 + n_2)^2]}$. Here, $n_1$ and $n_2$ are the carrier densities, and $\mu_1$ and $\mu_2$ are the carrier mobilities for band 1 and band 2, as described below. We have substituted the carrier density calculated from the SdH oscillation, $n = (1/3\pi^2)(2eF/\hbar)^{3/2}$ and found one hole band (band 1) with carriers from the α and β pockets ($n_1 = n_\alpha + n_\beta = 1.79 \times 10^{19}$ cm$^{-3}$, where $n_\alpha = 1.132 \times 10^{19}$ cm$^{-3}$ and $n_\beta = 0.658 \times 10^{19}$ cm$^{-3}$, respectively), and one electron band (band 2) with carriers from the γ pocket, $n_2 = n_\gamma = -2.42 \times 10^{19}$ cm$^{-3}$, which can give the best fit of the Hall data, yielding $\mu_1 = 1164.5$ cm$^2$V$^{-1}$s$^{-1}$ and $\mu_2 = 1045.5$ cm$^2$V$^{-1}$s$^{-1}$, respectively. The mobility is very close to the threshold value for observation of SdH oscillations: $\mu B \approx 1$ for a magnetic field of 5 to 10T. Therefore, the band structure near the Fermi level can be represented as in Fig. 4(c), and the three pairs Fermi pockets are identified as one pair of electron pockets and two pairs of hole pockets. Moreover, the quantum oscillation signal can be qualitatively fit by the LK formula with two oscillation frequencies, one corresponding to the electron pockets, $F_e = F_3$ and the other one corresponding to the two pairs of hole pockets, $F_h \approx (F_1^2 + F_2^2)^{1/2}$, as shown by the solid red line in Fig. 3(d). Therefore, the band structure near the Fermi level for the WTe$_2$ measured in this work can be schematically shown in Fig. 4(c). The Fermi pockets are schematically represented in Fig. 4(d).

Our FFT analysis of SdH oscillations reveals three pairs of Fermi pockets in our samples, instead of the two pairs of pockets reported in previous works [1, 8]. Our observations on the multiple Fermi pockets agree with the multiple valence bands obtained by first-principles calculations [1, 9]. The fit of the Hall resistivity indicates that one of the three pairs of Fermi pockets consists of electron pockets, while the other two consist of hole pockets, and the carrier densities of electrons and holes are nearly perfectly compensated. Our work suggests that the electronic structure of WTe$_2$ could be even more complicated than the two pairs of electron and hole pockets of approximately the same size, but as long as the electrons and holes are compensated the non-saturating parabolic MR will still persist.

Note added: During the preparation of this manuscript we noted three works reporting the structure of the Fermi pockets[20-22]. Among them, Refs. 20 and 21 are consistent with our work, but Ref. 19 attributes the extra Femi pockets to a second set of electron and hole pockets that may form along the Z-U direction.

This work is supported by the Australian Research Council (ARC) through an ARC Discovery Project (DP130102956, X.L.W.) and an ARC Professorial Future Fellowship Project (FT130100778, X.L.W.)

*Corresponding author.
xiaolin@uow.edu.au

[1] M. N. Ali *et al.*, Nature **514**, 205 (2014).

- [2] F. Y. Yang, K. Liu, K. Hong, D. H. Reich, P. C. Searson, and C. L. Chien, Science **284**, 1335 (1999).
- [3] Y. Kopelevich, J. Torres, R. da Silva, F. Mrowka, H. Kempa, and P. Esquinazi, Phys. Rev. Lett. **90**, 156402 (2003).
- [4] B. Fauqué, B. Vignolle, C. Proust, J.-P. Issi, and K. Behnia, New J. Phys. **11**, 113012 (2009).
- [5] Y. Zhao *et al.*, preprint at http://arxiv.org/abs/1502.04465 (2015).
- [6] X.-C. Pan *et al.*, Preprint at http://arxiv.org/abs/1501.07394 (2015).
- [7] D. Kang *et al.*, Preprint at http://arxiv.org/abs/1502.00493 (2015).
- [8] I. Pletikosić, M. N. Ali, A. V. Fedorov, R. J. Cava, and T. Valla, Phys. Rev. Lett. **113**, 216601 (2014).
- [9] H. Y. Lv, W. J. Lu, D. F. Shao, Y. Liu, S. G. Tan, and Y. P. Sun, Preprint at http://arxiv.org/abs/1412.8335 (2015).
- [10] D. Shoenberg, *Magnetic oscillations in metals* (Cambridge University Press, Cambridge, 1984).
- [11] S. E. Sebastian, N. Harrison, F. F. Balakirev, M. M. Altarawneh, P. A. Goddard, R. Liang, D. A. Bonn, W. N. Hardy, and G. G. Lonzarich, Nature **511**, 61 (2014).
- [12] D.-X. Qu, Y. S. Hor, J. Xiong, R. J. Cava, and N. P. Ong, Science **329**, 821 (2010).
- [13] J. G. Analytis, R. D. McDonald, S. C. Riggs, J.-H. Chu, G. S. Boebinger, and I. R. Fisher, Nature Phys. **6**, 960 (2010).
- [14] M. Veldhorst *et al.*, Nature Mater. **11**, 417 (2012).
- [15] F.-X. Xiang, X.-L. Wang, and S.-X. Dou, Preprint at http://arxiv.org/abs/1404.7572 (2014).
- [16] H. Murakawa, M. S. Bahramy, M. Tokunaga, Y. Kohama, C. Bell, Y. Kaneko, N. Nagaosa, H. Y. Hwang, and Y. Tokura, Science **342**, 1490 (2013).
- [17] F.-X. Xiang, X.-L. Wang, M. Veldhorst, S.-X. Dou, and M. S. Fuhrer, Preprint at http://arxiv.org/abs/1501.03240 (2015).
- [18] T. Liang, Q. Gibson, M. N. Ali, M. Liu, R. J. Cava, and N. P. Ong, Nature Mater. **14**, 280 (2015).
- [19] L. P. He, X. C. Hong, J. K. Dong, J. Pan, Z. Zhang, J. Zhang, and S. Y. Li, Phys. Rev. Lett. **113**, 246402 (2014).
- [20] P. L. Cai *et al.*, Preprint at http://arxiv.org/abs/1412.8298 (2014).
- [21] Z. Zhu, X. Lin, J. Liu, B. Fauque, Q. Tao, C. Yang, Y. Shi, and K. Behnia, preprint at http://arxiv.org/abs/1502.07797 (2015).
- [22] J. Jiang *et al.*, Preprint at http://arxiv.org/abs/1503.01422 (2015).